\documentclass{jps-cp}
\usepackage{txfonts} %Please comment out this line unless the txfonts package is availabe in your LaTeX system.
\usepackage{graphicx}

\title{SPD - the Spin Physics Project with
Polarized Proton and Deuteron
Beams at the NICA Collider}

\author{Alexey \textsc{Guskov}$^{1}$ on behalf of the SPD working group}

\inst{$^{1}$Joint Institute for Nuclear Research, Joliot-Curie 6, Dubna, 141980, Russia }

\email{avg@jinr.ru}

\recdate{}

\abst{The SPD experiment at the future NICA collider at JINR (Dubna, Russia) aims to investigate the nucleon spin structure and  polarization phenomena in collisions of longitudinally and transversely polarized protons and deuterons at $\sqrt{s}$ up to 27 GeV and luminosity up to 10$^{32}$ cm$^{-2}$ s$^{-1}$. Measurement of asymmetries in the Drell-Yan pairs, charmonium and prompt photon production can provide an access to the full set of leading twist TMD PDFs in nucleons. The experimental setup is planned as a universal 4$\pi$ detector for a wide range of physics tasks.
}

\kword{nucleon spin structure, TMP PDFs, SPD, NICA}

\begin{document}
\maketitle

\section{Introduction}
Since the famous ”spin crisis” that began in 1987, the problem of the nucleon spin structure remains one
 of the most intriguing puzzles in the contemporary high-energy physics. The central component of this
 problem, attracting for many years enormous both theoretical and experimental efforts, is the problem
 how the spin of the nucleon is built up of spins and orbital angular momenta of its constituents. The searches
 brought up a concept of the parton distribution functions of the nucleon, at first
 only two were proposed: $f_1$ for unpolarized and $g_1$ for polarized nucleons. Now we know that there must
 be of about 50 different parton distribution functions for a complete description of the nucleon structure.
 The most general of them is the Wigner function containing all
full information about the longitudinal and transverse spin and momentum partonic degrees of freedom.
The averaging over some variables or putting some other to particular (zero) values leads to more simple
 and more easily measurable description of fundamental particles, quarks and gluons, inside hadrons.
The important stage of this reduced description is the set of Transverse-Momentum-Dependent
Parton Distributions (TMD PDFs) and the reduction process finishes when the most familiar collinear parton
distributions appear.
 
NICA (Nuclotron-based Ion Collider fAсility) is a new accelerator complex designed at the Joint Institute
 for Nuclear Research (Dubna, Russia) to study properties of baryonic matter. In the first interaction point of the new collider 
 the MultiPurpose Detector  (MPD) 
 intends to study properties of hot dense nuclear matter
  in heavy ions collisions. The Spin Physics Detector (SPD) in the second interaction point is planned for
  investigation of the nucleon spin structure in collisions of longitudinally and transversely polarized protons and deuterons
   at $\sqrt{s}$ up to 27 GeV and luminosity up to 10$^{32}$ $s^{-1}cm^{-2}$. The measurement of the 
   TMD PDFs for  quarks and gluons using such reactions as the Drell-Yan process, charmonia and prompt photon
   production is the main goal of the experiment.
   The basic concept of the
   SPD project is presented below.

\section{SPD physics programme}
\subsection{Drell-Yan process}
Measurement of the azimuthal asymmetries in unpolarized and polarized Drell-Yan process provides access 
to the next TMD PDFs:

\begin{itemize}
\item transversity:  $A_{UT}^{\sin(\phi+\phi_S)}$, 
transverse polarization of quarks in  the transversely polarized nucleon;
\item Sivers function: $A_{UT}^{\sin(\phi-\phi_S)}$,
correlation of the transverse momentum of non-polarized quark with the nucleon transverse
polarization;
\item pretzelosity:  $A_{UT}^{\sin(3\phi-\phi_S)}$,
distribution of the transverse momentum of quarks in the transversely 
polarized nucleon;
\item Boer-Mulder function: $A_{UU}^{\cos(2\phi_h)}$,
distribution of the quark transverse momentum in the non-polarized nucleon;
\item worm-gear functions:  $A_{UL}^{\cos(2\phi_h)}$,
correlation between the longitudinal polarization of the nucleon (longitudinal spin)
and the transverse momentum of quarks.
\end{itemize}
All mentioned PDFs are proposed to be measured with 
SPD \cite{Savin:2014sva}. 
At the twist-3 approximation there are already 16 TMD PDFs 
containing the information on the nucleon structure. 
They have no definite physics interpretation yet.

Two approaches are proposed to be used for TMD PDFs  extraction \cite{Savin:2014sva}.
The first one  is based on the Fourier analysis \cite{dy1,dy2,dy3,dy4,dy5,dy6,dy7},
while the second one is the extraction of PDFs via integrated/weighted azimuthal asymmetries
\cite{dy8,dy9,dy10,dy11}. 

\subsection{Prompt-photon production}
Prompt-photon production in hadron collisions, where the gluon Compton scattering dominates as a hard process, is the most direct and proven way to access the gluon structure of nucleons. 

The measurement of single transverse spin asymmetry $A^{\gamma}_N$ in prompt-photon production at high $p_T$ in transversely polarized p-p and d-d collisions could provide information on the gluon Sivers function which is almost unknown at the moment \cite{pp1}. The authors of the work \cite{pp3} pointed out that $A^{\gamma}_N$ at large positive $x_F$ is dominated by quark-gluon correlations while at large negative $x_F$ it is dominated by pure gluon-gluon correlations as it was concluded in \cite{pp4}.
 The study of prompt-photon production at large transverse momentum with longitudinally polarized proton beams could provide the access to gluon polarization $\Delta{g}$, which is known from previous measurements to be of about zero with large uncertainties, via the measurement of longitudinal double spin asymmetry $A^{\gamma}_{LL}$ \cite{pp5}.

Unpolarized measurements of the differential cross section of prompt-photon production 
in proton-proton collisions at SPD is also interesting since there is a significant disagreement between the theoretical expectations and the experimental results in the fixed-target experiments at $\sqrt{s}\sim 20$ GeV  \cite{pp6}. There is no such strong disagreement for the high-energy collider results.

\subsection{Charmonia production}
The production of lepton pairs in the $q\bar{q}$ annihilation processes, via the $J/\psi$ production and its subsequent decay is analogous to the Drell-Yan production mechanism. The analogy is correct if the $J/\psi$ interactions with quarks and leptons are of the vector type. This analogy is known as the duality model \cite{jpsi1,jpsi2}. For the TMD PDFs
studies, the duality model can predict \cite{jpsi3} a similar behavior of the azimuthal asymmetries in the
lepton pair's production via Drell-Yan  and via the $J/\psi$ leptonic decay This similarity stems from the idea of the duality model to replace the coupling $e_q^2$ by the $J/\psi$ vector coupling with $q\bar{q}$ $(g_q^V)^2$.
 The vector couplings are expected to be the
 same for $u$ and $d$ quarks \cite{jpsi1} and cancel out in the azimuthal asymmetries for large values of $x_1$ or $x_2$. 
 For instance  the Sivers asymmetry for Drell-Yan process could be compared with the corresponding asymmetry for  $J/\psi$ events 
 at different colliding beam energies.
 
Unpolarized $J/\psi$ and $\chi_{cJ}$ production at the SPD energies is also interesting from the point of 
 separation of quark-antiquark annihilation and gluon-gluon fusion contributions. Production of
 other charmonia including exotic states like $X(3872)$ could be performed in parallel.\\
 
Phenomena with high-$p_t$ hadrons (polarized and unpolarized case), multiquark correlation effects, study of generalised parton distributions (GPDs) in the reaction of deeply virtual meson production ($pp\to pp M$) are also the parts of the SPD physics programme. 
\begin{figure}[h!]
\centering
\includegraphics{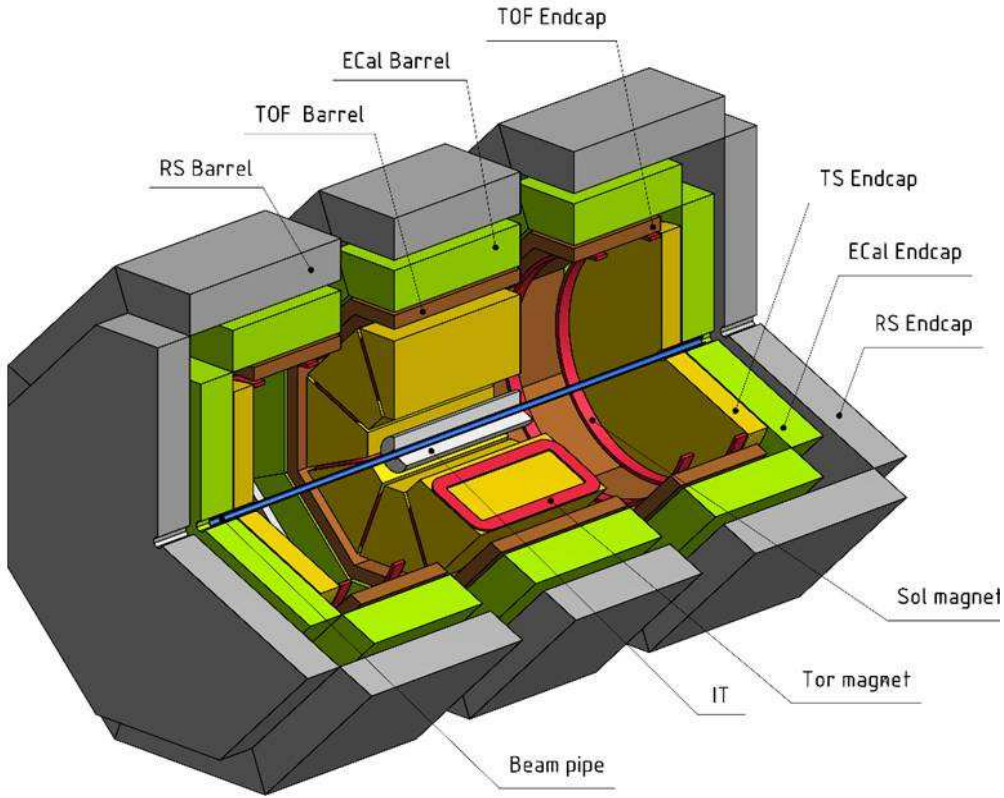}
\caption{Conceptual layout of the SPD setup \cite{CDR}. The main detector elements are shown.}
\label{f1}
\end{figure}
\section{The SPD setup}

A conceptual layout of the SPD setup, that is planned as a universal 4$\pi$ detector, is presented in Fig. \ref{f1} \cite{CDR}.
The detector consists of the three parts: barrel and two end-caps. Each part has individual magnetic system. 
The toroidal field should provide tracking capability in the barrel part while the end-caps will be equipped by the 
solenoidal coils. Such configuration of the magnetic system should minimize the magnetic field in the beam interaction region 
that is especially important for transversely polarized beams and provide the field integral of about 1 T$\cdot$m for measurement 
of the momentum of secondary charged particles.
The planned length of the detector along
 the beam axis is 9.2 m while the diameter is 3.4 m. 

Tracking system (TS) of the SPD setup consists of the silicon-based inner tracker (IT) in the central part of the detector surrounded by 
the main tracker using gas-filled drift straw-tubes as the basic detection element.
Time-of-flight system (TOF) will provide identification of secondary hadrons in the wide kinematic range. 
Low material budget of the central part of the setup is important for transparency of the detector for photons and minimization of the
multiple scattering effects.
The shashlyk-type electromagnetic calorimeter  (ECal) 
is responsible for the photon reconstruction. It will be the main element for the prompt phonon part of the physics programme. 
 The range (muon) system (RS), the main detector for the Drell-Yan and charmonia part of the programme, should perform the advanced muon identification
via muonic pattern recognition and further matching of the track segments to the tracks in the inner part of the detector.

\section{Summary}
The measurement of azimuthal asymmetries in the Drell-Yan process in collisions of unpolarized, longitudinally and transversally polarized proton and deuteron beams are planned to be performed at the NICA collider that is under construction at JINR. These measurements will provide access to all leading twist TMD PDFs of quarks and anti-quarks in nucleons. The measurements of asymmetries in production of charmonia and prompt photons, which supply complimentary information on the nucleon structure, will be performed in parallel. The Spin Physics Detector is planned to be a universal detector for a wide range of physics tasks. More detailed information about the SPD project could be found in the Letter of Intent \cite{Savin:2014sva} and the preliminary version of the Conceptual Design Report \cite{CDR}.


\begin{thebibliography}{9}
\bibitem{Savin:2014sva}
  I.~A.~Savin {\it et al.},
  EPJ Web Conf.\  \textbf{85} 02039  (2015). 
\bibitem{dy1} S. Arnold, A. Metz, M. Schlegel,  Phys. Rev. \textbf{D79}  034005 (2009).
\bibitem{dy2} L. P. Gamberg, D. S. Hwang, A. Metz, M. Schlegel, Phys. Lett. \textbf{B639} 508  (2006).
\bibitem{dy3} A. Bacchetta, D. Boer, M. Diehl, P. J. Mulders, JHEP \textbf{08}  023 (2008).
\bibitem{dy4} M. Anselmino, M. Boglione, U. D’Alesio, S. Melis, F. Murgia, A. Prokudin, Phys. Rev.  \textbf{D88} 054023  (2013).
\bibitem{dy5} D. Boer, L. Gamberg, B. Musch, A. Prokudin, JHEP \textbf{10}  021 (2011).
\bibitem{dy6} P. J. Mulders, R. D. Tangerman, Nucl. Phys. \textbf{B461} 197 (1996).
\bibitem{dy7} A. Bacchetta, M. Diehl, K. Goeke, A. Metz, P. J. Mulders, M. Schlegel, JHEP \textbf{02} 093 (2007). 
\bibitem{dy8} A. Sissakian, O. Shevchenko, O. Ivanov, JETP Lett \textbf{86} 863  (2007).
\bibitem{dy9} A. N. Sissakian, O. Yu. Shevchenko, A. P. Nagaytsev, O. N. Ivanov, Phys. Rev. \textbf{D72} 054027 (2005).
\bibitem{dy10} A. Sissakian, O. Shevchenko, A. Nagaytsev, O. Denisov, O. Ivanov,  Eur. Phys. J.  \textbf{C46}  147 (2006).
\bibitem{dy11} A. N. Sissakian, O. Yu. Shevchenko, A. P. Nagaitsev, O. N. Ivanov, Phys. Part. Nucl. \textbf{41}  64 (2010).
\bibitem{pp1} D. Boer, C. Lorce, C. Pisano, J. Zhou, Adv. High Energy Phys \textbf{2015} 371396 (2015).
%\bibitem{pp2} L. Gamberg, Z.-B. Kang,  Phys. Lett. \textbf{B718} 181 (2012).
\bibitem{pp3} J.-W. Qiu, G. F. Sterman,  Nucl. Phys. \textbf{B378}  52 (1992). 
\bibitem{pp4} X.-D. Ji, Phys. Lett. \textbf{B289} 137 (1992). 
\bibitem{pp5} H.-Y. Cheng, S.-N. Lai, Phys. Rev. \textbf{D41} 91  (1990).
\bibitem{pp6} P. Aurenche, M. Fontannaz, J.-P. Guillet, E. Pilon, M. Werlen, Phys. Rev. \textbf{D73} 094007 (2006).
\bibitem{jpsi1} M. Anselmino, V. Barone, A. Drago, N. N. Nikolaev,  Phys. Lett. \textbf{B594} 97 (2004).
\bibitem{jpsi2} E. Leader, E.  Predazzi, An introduction to gauge theories and the new physics, Cambridge Univ. Press. (1982).
\bibitem{jpsi3} V. Barone, Z. Lu, B.-Q. Ma,  Eur. Phys. J. \textbf{C49}  967 (2007).
\bibitem{CDR} Conceptual and technical design of the Spin Physics Detector (SPD) at the NICA collider \mbox{https://indico.jinr.ru/getFile.py/access?resId=11\&materialId=3\&confId=718} (2018).

\end{thebibliography}
\end{document}